\documentclass[fleqn,usenatbib]{mnras}

\usepackage{newtxtext,newtxmath}
\usepackage[T1]{fontenc}
\usepackage{ae,aecompl}
\usepackage{url}
\usepackage{graphicx}
\usepackage{xspace}
\usepackage{amsfonts}
\usepackage{pifont}
\usepackage{textcomp}
\usepackage{float}
\usepackage{subfig}
\usepackage{multirow}
\usepackage{adjustbox}

\title[QPPs from low mass stars]{Doubling of minute-long Quasi-Periodic Pulsations from super-flares on a low mass star}

\author[Doyle et al.]{J. Gerry Doyle,$^{1}$\thanks{E-mail: gerry.doyle@armagh.ac.uk}
Puji Irawati,$^{2}$ Dmitrii Y. Kolotkov,$^{3}$ Gavin Ramsay,$^{1}$ Nived Vilangot Nhalil,$^{1}$ 
\newauthor
Vik S. Dhillon,$^{4,5}$ Tom R. Marsh$^{6}$, Ram Kesh Yadav$^{2}$\\
$^{1}$ Armagh Observatory \& Planetarium, College Hill, Armagh, BT61 9DG, N. Ireland\\
$^{2}$ National Astronomical Research Institute, 260 Moo 4 Don Kaeo, Mae Rim District, Chiang Mai 50180, Thailand\\
$^{3}$ Centre for Fusion, Space and Astrophysics, Department of Physics, University of Warwick, Coventry, CV4 7AL, UK\\
$^{4}$ Department of Physics and Astronomy, University of Sheffield, Sheffield S3 7RH, UK\\
$^{5}$Instituto de Astrof\'{i}sica de Canarias, E-38205 La Laguna, Tenerife, Spain\\
$^{6}$ Department of Physics, University of Warwick, Coventry CV4 7AL, UK
}

\begin{document}
\outer\def\gtae {$\buildrel {\lower3pt\hbox{$>$}} \over 
{\lower2pt\hbox{$\sim$}} $}
\outer\def\ltae {$\buildrel {\lower3pt\hbox{$<$}} \over 
{\lower2pt\hbox{$\sim$}} $}
\newcommand{\Msun}{$M_{\odot}$}
\newcommand{\lsun}{$L_{\odot}$}
\newcommand{\Rsun}{$R_{\odot}$}
\newcommand{\solar}{${\odot}$}
\newcommand{\kep}{\sl Kepler}
\newcommand{\ktwo}{\sl K2}
\newcommand{\tess}{\sl TESS}
\newcommand{\swift}{\it Swift}
\newcommand{\Porb}{P_{\rm orb}}
\newcommand{\nuorb}{\nu_{\rm orb}}
\newcommand{\eplus}{\epsilon_+}
\newcommand{\eminus}{\epsilon_-}
\newcommand{\cd}{{\rm\ c\ d^{-1}}}
\newcommand{\MdotL}{\dot M_{\rm L1}}
\newcommand{\Mdot}{$\dot M$}
\newcommand{\Mdotsolar}{\dot{M_{\odot}} yr$^{-1}$}
\newcommand{\Ldisk}{L_{\rm disk}}
\newcommand{\src}{KIC 9202990}
\newcommand{\ergscm} {erg s$^{-1}$ cm$^{-2}$}
\newcommand{\rchi}{$\chi^{2}_{\nu}$}
\newcommand{\chisq}{$\chi^{2}$}
\newcommand{\pcmsq} {cm$^{-2}$}

\providecommand{\lum}{\ensuremath{{\cal L}}}
\providecommand{\mg}{\ensuremath{M_{\rm G}}}
\providecommand{\bcg}{\ensuremath{BC_{\rm G}}}
\providecommand{\mbolsun}{\ensuremath{M_{{\rm bol}{\odot}}}}
\providecommand{\teff}{\ensuremath{T_{\rm eff}}}

\maketitle
\begin{abstract}
Using the ULTRASPEC instrument mounted on the 2.4-m Thai National Telescope, we observed two large flares, each with a total energy close to 10$^{34}$ erg with sub-second cadence. A combination of a wavelet analysis, a Fourier transform plus an empirical mode decomposition, reveals quasi-period pulsations (QPP) which exhibit an apparent doubling of the oscillation period. Both events showed oscillations of a few minutes over a interval of several minutes, and despite the availability of sub-second cadence, there was no evidence of sub-minute oscillations. The doubling of the QPP periods and shorter lifetime of shorter-period QPP modes strongly favour resonant dynamics of magnetohydrodynamic waves in a coronal loop. We estimate loop lengths to be 0.2--0.7 $R_\star$, in agreement with a typical length of solar coronal loops. These observations presents rare and compelling evidence for the presence of compact plasma loops in a stellar corona.

\end{abstract}

\begin{keywords}
  stars: activity -- stars: flare -- stars: low-mass
  -- stars: late-type -- stars: rotation -- stars: oscillations
\end{keywords}

\section{Introduction}

M-dwarf stars are the most numerous in the solar neighborhood constituting $\sim$80\% of the galactic stellar 
population. They are characterized by a very low mass range (0.075 – 0.6 M$\sun$) and low effective temperatures 
2500 (M9.5) - 3700 K (M0) but with exceptionally long stellar lifetimes. Their close proximity to Earth, vast number density, stellar age, their small size relative to solar-like stars hence deeper transits, make them attractive targets for observational facilities dedicated to the search for exo-planetary systems \citep{Shields2016}. While M dwarfs are advantageous targets in the search for exoplanets, a potential complication lies is their intrinsic variability. In particular, M dwarfs experience strong magnetic activity on scales orders of magnitude greater than the Sun resulting in intense flaring. They frequently have “super-flares”, which have a total bolometric energy $10^4$ greater than the largest solar flares, \citep{Yang2017, Paudel2019}.

To truly understand the link between stellar and solar flares, we must first create a solid observational link between the physical processes occurring in each case. The largest Earth-directed solar flare on record was the 1859 “Carrington flare”, which brought a halt to the telegraph network across much of Europe and North America. This flare was orders of magnitude less energetic than the stellar super-flares observed on M dwarfs, e.g. see \cite{Ramsay2021} who reported flare energy in excess of 10$^{36}$ ergs. Studying stellar flares is vital for understanding the mechanisms responsible for magnetic fields in stars, and the physical processes responsible for flares, and space weather.

Quasi-periodic pulsations (QPPs), with periods between fractions of a second to tens of minutes, are a feature of solar flares that also exist in stellar flares, although the shortest periods observed in stellar flares is 20 sec., e.g. \citet{Welsh2006}, \cite{Mathioudakis2006} and \citet{Doyle2018}. Solar QPPs provide information concerning properties of the associated active region, e.g. \citet{Nakariakov2009}. QPPs appear to be a common feature of solar flares \citep{Simoes2015}, and their dominant periods and decay times provide information about the physical properties of the star's local active region such as the Alfv\'en and sound speeds, e.g. \citet{Nakariakov2011}. In white-light stellar flare data, QPPs typically appear as modulations in the decay
phase of large flares or super-flares. QPPs have been suggested as a potential means to connect the physics of solar and stellar flares. 

Different plausible explanations for these periodic signatures include repetitive reconnection and
magnetohydrodynamic oscillations, analogous to those observed in solar flares. Reviews by 
\citet{McLaughlin2018}, \citet{Kupriyanova2020} and \cite{Kol2021} discuss a range of possible QPPs mechanisms for solar and stellar flares. Due to the multitude of possible mechanisms, it is very difficult to select an appropriate one, however, nature sometimes comes to the rescue via showing a series of period harmonics which can guide us.   
The above works imply a link between stellar flares and stellar magnetic activity, and that the same physical processes are involved in solar and stellar flares. A dedicated survey, along with a combined consideration of solar and stellar flares may
allow for scaling laws to be established, akin to those proposed by \cite{Mathioudakis2006} and 
\citet{Aschwanden2008}. Additional samples of QPPs in large flares at high cadence are desperately needed to 
improve our ability to disentangle competing physical mechanisms. \citet{Ramsay2021} performed a search for flares and QPPs from low-mass M-dwarf stars using Transient Exoplanet Survey Satellite (TESS) two-minute cadence data. These authors determined the length and magnetic-field strengths of the flare coronal loops using the period of the QPPs and various assumptions about the origin of the QPPs.

Flares have been observed on a diverse range of stars for many years. However, interest in stellar
flares has recently increased because of observations on solar-like stars of flares many orders of magnitude larger than even the largest Earth-directed flare ever observed on our own Sun, e.g. \citet{Maehara2012}, \citet{Notsu2019}. Kepler and now TESS have provided us the means for mass exploration of the flaring rates on thousands of M dwarfs. Due to it’s 2 min and 20 sec cadence rates, this facility is only suitable for the detection of QPPs of several minutes. \cite{Maehara2021} reported on 145 TESS flares, some with ground-based H$\alpha$ observations, but there was no discussion of QPPs.

Here, we use high-speed photometry data gained via the 2m-class Thai National Telescope (TNT) to explore the active flare star, YZ CMi. Prior to TESS, the most extension flare monitoring of this star was presented by \citet{Ishida1991} in the U-band;  reporting on over 300 hrs of U-band monitoring based on data obtained between 1972 and 1987.
In Section 2, we discuss the observations and their reduction; Section 3 looks at the different methods used to search for QPPs. 
Finally, in Section 4, we explore the implications of the findings.

\begin{table}
\caption{Observation log for YZ CMi flare monitoring from the Thai National Telescope}

\begin{tabular}{llllll}
\hline
Date 	    & Target & Sampling(sec) &	Filter	 & 	Seeing(") &	Duration\\
21 Jan 2018 & YZ CMi & 0.5063	  & $u^,$	     &	1.6-2.4	  &	31 min\\ 
23 Jan 2018 & YZ CMi & 0.5305	  & $u^,$	     &	1.8-3.2	  &	310 min\\    
19 Feb 2018 & YZ CMi & 0.6562	  & $u^,$	     &	1.4-3.5	  &	210 min\\
22 Mar 2018 & YZ CMi & 0.2509	  & $g^,$   	 &	1.7-3.2	  &	99 min\\ 
06 Nov 2018 & YZ CMi & 0.5631	  & $u^,$	     &	1.8-2.4	  &	67 min\\ 
07 Nov 2018 & YZ CMi & 0.6038	  & $u^,$	     &	2.0-4.0	  &	270 min\\ 
07 Apr 2019 & YZ CMi & 0.2762	  & $g^,$  	     &	1.3-2.2	  &	96 min\\ 
\hline
\end{tabular}
\label{observing}
\end{table}

\section{Observations from the TNT}

We used the ULTRASPEC instrument \citep{Dhillon2014} mounted on the 2.4-m Thai National Telescope (TNT),
located on Doi Inthanon in Thailand. Two filters were used for the flare monitoring; $u^,$ at a central 
wavelength of 355.7 nm and width 59.9 nm plus a $g^,$ filter at a central wavelength of 482.5 nm and width 
of 137.9 nm. The active dMe star, YZ CMi, was observed in 2018 with a cadence ranging from 0.25 to 0.65 sec, while 
in 2019, YZ CMi was observed with a cadence of 0.27 sec, see Table \ref{observing}. The ULTRASPEC dead time between exposures is only 15 msec. Several flares were detected, here we only discuss the two largest flares.

\begin{figure}
\centering
\includegraphics[width=0.5\textwidth]{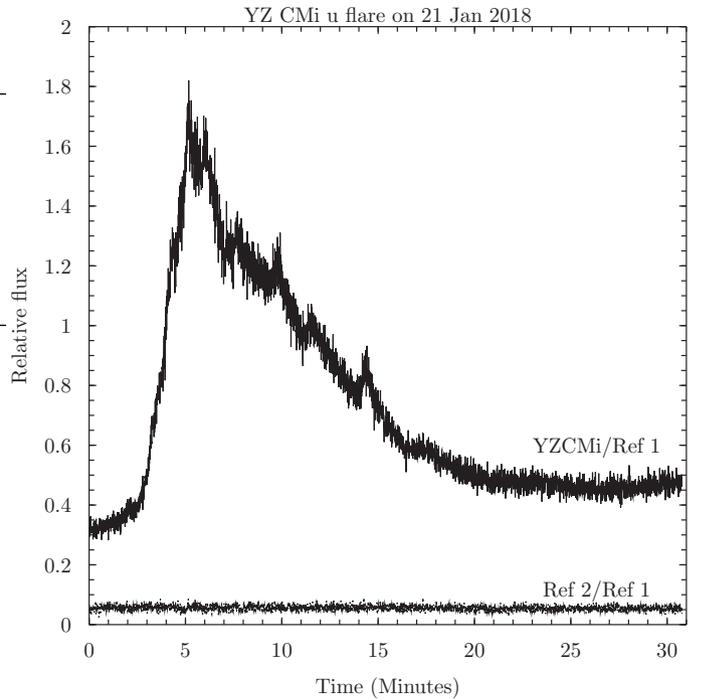}
\vspace*{0.5cm}
\caption{The YZ CMi light-curve in the u$^,$ band from 21 Jan 2018 divided by the flux of one of the reference stars. The bottom line shows the ratio of the two reference stars.}
\label{yzcmi_1}
\end{figure}

\begin{figure}
  \vspace*{-0.5cm}
  \centering
\includegraphics[width=0.5\textwidth]{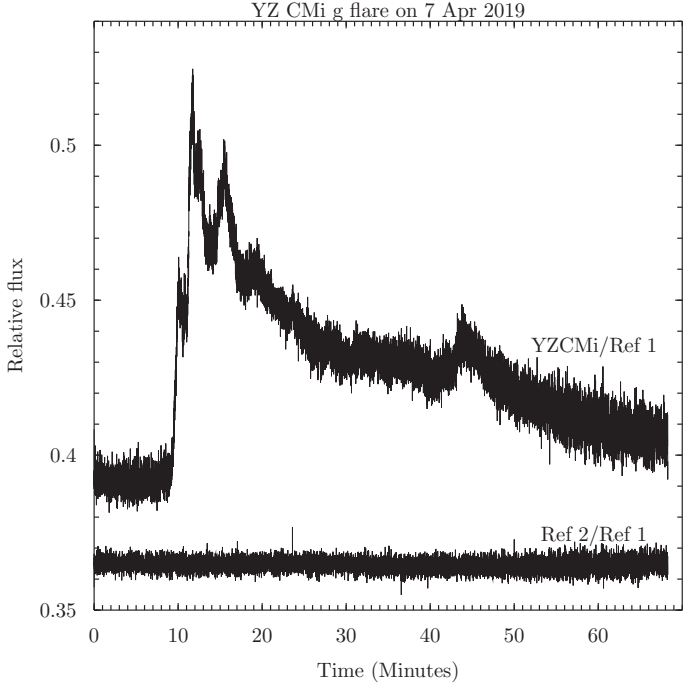}
\vspace*{-0.2cm}
\caption{As in Figure~\ref{yzcmi_1} but for the YZ CMi flare taken in the g$^,$-band on 7 Apr 2019.}
\label{yzcmi_2}
\end{figure}

All of the photometric data in the $u^,$-band were acquired using the avalanche output with a two-window mode. 
Filter $g^,$ was obtained in normal output when the sky conditions were not favourable to reach SNR>30 in the $u^,$-filter. Each of our observing runs had a duration between 30-300 minutes, the largest flares were observed 
on 21 Jan 2018 and 07 Apr 2019. The data were then processed using the ULTRACAM pipeline \citep{Dhillon2007} where we applied bias correction and flat fielding to all of our science images. We performed aperture photometry to get 
the flux of our target star with variable aperture to match the seeing conditions. Two reference stars were also recorded simultaneously on the second window to monitor the transparency during the observation. We converted the counts into magnitudes using the ULTRASPEC zero-point magnitudes, then we compute the flux for YZ CMi in each band-pass. This work assumes a black-body of 8,000 K. See Figures~\ref{yzcmi_1} and \ref{yzcmi_2} for the light-curves of the two larger flares. Also shown in these figure is the ratio of the flux from the two reference stars, indicating no features were due to changes in seeing or transparency.

  \begin{figure}
  \centering
\includegraphics[width=0.5\textwidth]{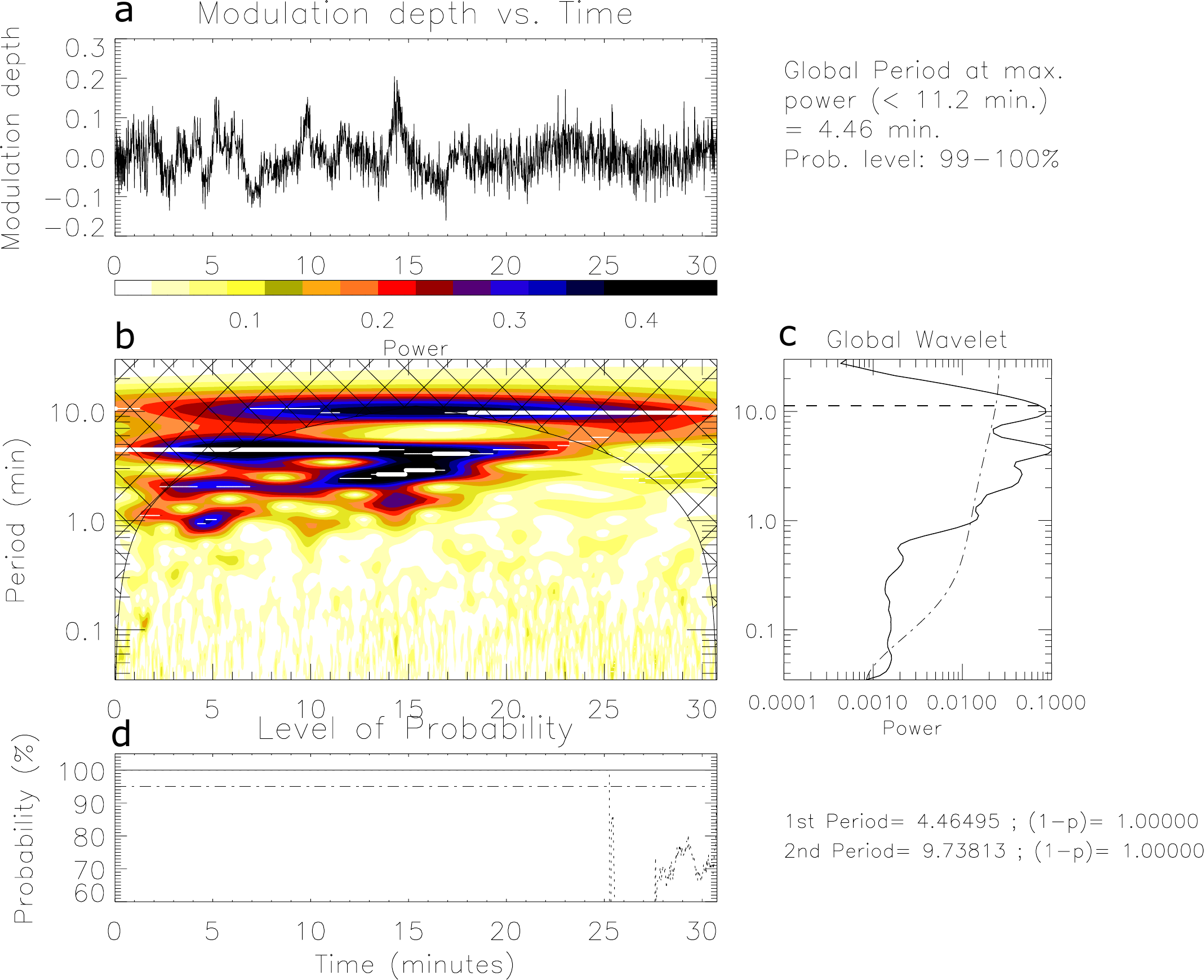}
\caption{From top to bottom; panel (a) shows the modulation depth for the YZ CMi flare of 21 Jan 2018, panel (b) shows the resulting wavelet spectrum. In panel (c) we show the global wavelet spectrum (which is the average of the local wavelet spectrum (panel b) over time) and the various derived periods; the dashed-dotted line shows the 95\% confidence level. Panel (d) shows the probability level for the first maximum which remains at 100\% all of the time. But for the second maximum, the probability reduces drastically after 25 minutes. The dashed-dotted line shows the 95\% confidence level.} 
\label{wavelet_2018}
\end{figure}

  \begin{figure}
  \hspace*{-6cm}
  
\includegraphics[width=0.5\textwidth]{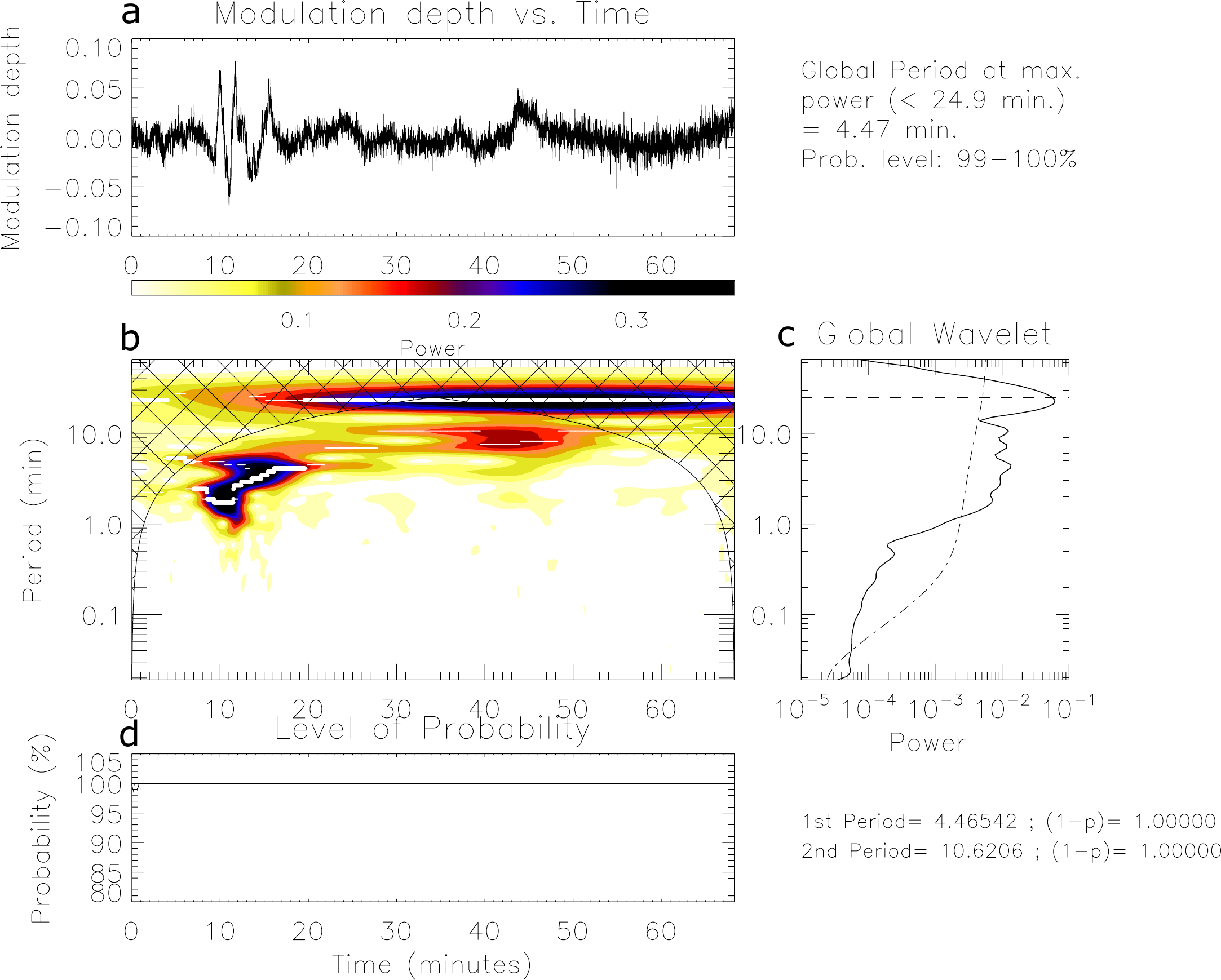}
\caption{Same as in Figure \ref{wavelet_2018} but for the  flare of 07 Apr 2019.} 
\label{wavelet_2019}
\end{figure}

\section{Results}
\label{tess}
\subsection{TNT flare energetics}

To calculate the quiescent luminosity, we used the Gaia DR2 distance from \citet{Bailer-Jones2018} where they estimated YZ CMi distance as d=5.986 pc. For the $u^,$ band quiescent luminosity we derived $8.11 \pm 0.69 \times 10^{29}$ erg s$^{-1}$ (based on data from 23 Jan 2018), then using the flare's equivalent duration 
(see \cite{Ramsay2013} for further details) gives a flare energy of at least $8.80 \pm 0.014 \times 10^{33}$ erg for the 21 Jan 2018 event. For the $g^,$ band, the quiescent luminosity is $1.26 \pm 0.048 \times 10^{30}$ erg s$^{-1}$, which using the flare's equivalent duration gives a flare energy of at least $5.76 \pm 0.002 \times 10^{33}$ erg for the 07 Apr 2019 event. The formal uncertainties of the flare energy are very small as this reflects the fluxes uncertainties from the photometry, although we should note that these are lower limits as we do not observe the end of either flare. Furthermore, if the assumed black-body temperature is too low, then this could lead to an under estimate in the derived $u^,$ and $g^,$-band energy. \citet{Howard2020} showed that if super-flares are hotter than the above assumed black-body temperature of 8000 K, then the UV emission may be 10× higher than predicted from the optical. Furthermore, these authors show that the amount of time flares emit at temperatures above 14,000 K depends on energy; for example 43\% of the flares emit above 14,000 K, 23\% emit above 20,000 K and 5\% emit above 30,000 K. In addition, this is only the radiative energy from a small spectral region, \citet{Doyle1985} showed the total radiative emission is a factor of 14 times the U-band emission, which means that the energy of both YZ CMi flares are much greater than the largest solar flare which makes them super-flares, perhaps in excess of 10$^{35}$ erg. 

\subsection{QPPs determination}

Following the work of \citet{Broomhall2019}, the analysis of QPP signals was carried out using three methods: (i) a wavelet analysis, (ii) a Fourier transform, plus (iii) a method of empirical mode decomposition (EMD) on the modulation depth. In the above we used a self-consistent de-trending method and assessment of the statistical significance of the revealed intrinsic oscillatory modes in comparison with the background coloured noise, see  \cite{2016A&A...592A.153K} for further details. For the wavelet, we use the prescription given in \citet{Torrence1998} and the accompanying software.  We use a Morlet wavelet transform and a 95\% (2$\sigma$) significance level.
Figures \ref{wavelet_2018} \& \ref{wavelet_2019} show the result of wavelet analysis for the YZ CMi flare of 21 Jan 2018 and 07 Apr 2019. The wavelet analysis was performed on the flare's modulation depth which helps when the period strength changes during the flare. We derive the modulation depth by subtracting the trend from the original light curve and then dividing it by the trend. The flare trend for both events were obtained through a combination of fitting the observed light-curves with a prescribed theoretical model and EMD, as described in Section~3 of \citet{Ramsay2021}. As a model to fit, we used an asymmetric function with a rapid Gaussian rise and gradual exponential decay, mimicking a typical flare shape, superimposed onto a low-frequency harmonic component to account for possible background variations \citep[see e.g.][]{Kul2021}.

More in-depth details on the Fourier and EMD methods can be found in \citet{Ramsay2021}. The modulation depth plot is shown in panel (a) and the corresponding local power spectrum is shown in panel (b). Panel (c) presents the time-averaged global wavelet spectrum. The dashed-dotted line marks the 95\% confidence level. In panel (d), we plot the significance level derived from the randomisation procedure, which is independent of the underlying noise model.

For the 2018 flare, the Fourier and wavelets methods give consistent results with periods at 1.1, 2.2 \& 4.5 min with a clear doubling of the oscillation period (see Figure~\ref{wavelet_2018}, for the illustration of the wavelet analysis). All oscillations are observed to start at approximately the same time; 3 minutes of elapsed time in Figure~\ref{wavelet_2018} (i.e. during the rise phase of the flare), but have apparently different lifetimes: $\sim$4 minutes for the 1-min oscillation, $\sim$12 minutes for 2-min oscillation, and $\sim$15 minutes for 4-min oscillation. The 10 min period is within the cone of influence and therefore can not be considered real. The EMD analysis only finds the two shorter periods.

For the 2019 flare, quasi-periodic behaviour during the impulsive phase is seen as a rather broad spot in the wavelet dynamic spectrum, with periods similar to the 2018 flare (Figure~\ref{wavelet_2019}). A similar enhancement of the oscillation power in this range of periods is observed in the Fourier spectrum of the de-trended signal. The EMD analysis of this segment of the light-curve (from 6 to 22 minutes of the elapsed time in Figure~\ref{wavelet_2019}) has allowed us to distinguish two oscillatory modes with mean periods about 1.3 min and 2.8 min, which start approximately simultaneously and live for $\sim$5 minutes and $\sim$10 minutes, respectively. An apparent increase of the oscillation period seen in the time domain and in the wavelet spectrum can be attributed to a quicker decay of a shorter-period oscillation. The apparent 10 min period is perhaps due to a small flare seen at minute 45 in Figure~\ref{yzcmi_2}. A summary of the derived QPPs and flare energies are given in Table \ref{QPP}.

\begin{table}
\caption{Observed QPPs from the wavelet, EMD and Fourier analysis, plus the flare energy as seen in the $u'$ or $g'$ filters.}

\begin{tabular}{llll}
\hline
Object/Date              & Wavelet  & EMD \& Fourier     & E$_{band}$\\
                    & (minutes)       & (minutes)  & (erg)     \\
\hline
YZ CMi 21 Jan 2018  & 1.1, 2.2, 4.5   & 0.9, 2.0           & $8.80 \times 10^{33}$      \\
YZ CMi 07 Apr 2019   & 1.7 <--> 4.5   & 1.3, 2.8           & $5.76 \times 10^{33}$      \\
\hline
\end{tabular}
\label{QPP}
\end{table}

\section{DISCUSSION}

In the YZ CMi flare from Jan 2018, we observed at least three QPPs ranging from 1.1 to 4.5 min (Figure \ref{wavelet_2018}) while for the YZ CMi event from Apr 2019, we observed three periods of similar duration (Figure \ref{wavelet_2019}). In both flares, the observed QPP modes exhibit an apparent doubling of the oscillation period and quicker decay of shorter-period oscillations. For both events, the observed QPP are much shorter than YZ CMi rotation period of 2.774 days, based on TESS data over several rotational cycles, also see \cite{Maehara2021}, which allows us to attribute them to quasi-periodic dynamics in a localised flare-hosting active region.

There are several natural scenarios for the initial flare-caused impulsive perturbation of an active region to develop into a quasi-periodic response, which includes the effects of resonance in closed coronal plasma structures (acting as resonators), dispersion of a wave-guide, and nonlinearity/self-organisation \citep{McLaughlin2018, 2021SSRv..217...66Z}. The observed doubling of the QPP and shorter lifetime of shorter-period QPP modes strongly indicate in favour of resonant dynamics of magnetohydrodynamic waves in a coronal loop, for which the oscillation period is prescribed by the local plasma conditions (i.e. the Alfv\'en and sound speeds) and the loop length \citep{2021SSRv..217...34W, 2021SSRv..217...73N}. More specifically, fast- and slow-mode magnetoacoustic waves in coronal plasma structures are well known to be subject to a frequency-dependent damping by e.g. resonant absorption and thermal conduction \citep[e.g.][]{2002ApJ...577..475R, 2002ApJ...576L.153O, 2003A&A...408..755D}. Likewise, excitation of even and/or uneven parallel harmonics of fast- and slow-mode standing waves in a plasma loop is known to be highly sensitive to the location of the initial perturbation along the loop. For example, \citet{2004A&A...422..351T} and \citet{2005A&A...436..701S} theoretically demonstrated that the second parallel harmonic of a slow standing wave can be effectively excited if the impulsive energy release occurs near the apex of the loop. Observations of higher harmonics of fast-mode oscillations in coronal loops were also shown to be subject to the excitation mechanism and location of the initial displacement of the loop \citep[e.g.][]{Moortel2007, Sri2008, Yuan2016, Pascoe2017, Duck2019}. Thus, taking $c_\mathrm{s}=600$\,km\,s$^{-1}$ for the sound speed in a hot flaring loop, the Alfv\'en speed $c_\mathrm{A}=1200$\,km\,s$^{-1}$ \citep[e.g.][]{Mathioudakis2006}, and treating the observed QPP periods as characteristic acoustic or Alfv\'en transit times along the loop, we can estimate the corresponding loop lengths as 80--160\,Mm for the 2018 flare and 50--100\,Mm for the 2019 flare, i.e. 0.2--0.7 $R_\star$. These estimations agree with a typical length of solar coronal loops, and, are approximately an order of magnitude larger than those derived by \cite{Mathioudakis2006}. However, the above authors ruled out these small loop lengths suggesting instead that they may be due to a fast-MHD wave, with the modulation of the emission being due to the magnetic field. The present observation in the doubling of the QPP in both YZ CMi flares presents rare and compelling evidence for the presence of compact plasma loops in a stellar corona. 

\cite{Namekata2017} proposed a scaling law to estimate the loop length based on the correlation between the flare energy and flare duration. Using this relationship, the two flares reported here suggest a loop length of 2500-3000 Mm, which is a factor of $\sim$20-50 times larger than the loop lengths estimated using the observed QPPs. An intrinsic  assumption of the above method is that a single loop supplies the flare energy. Observations of solar flares from various space missions such as Hinode \citep{Kosugi2007} and the Solar Dynamic Observatory \citep{Pasnell2012} shows that most flares have tens of loops which expand into the corona, many releasing their energy via coronal rain and various dynamic process. Combining this scaling law with information on the loop length based on the QPPs suggest that these mega-flares have 20–50 loops.

As noted earlier, we did not find evidence of sub-minute and sub-second QPP, predicted, for example, by the oscillatory regime of the coalescence instability \citep{Tajima1987, Kol2016} or wave-particle interaction \citep[e.g.][]{Asch1987}, see also Table 1 in \citet{2021SSRv..217...66Z}. Although we have good agreement between the QPP periods and loop lengths between solar and stellar flares, we need to point out that estimations of the oscillation period, predicted by various QPP mechanisms, may differ significantly for physical conditions in solar flares and in powerful stellar flares. For example, \citet{Kol2021} showed that the QPP mechanism based on including additional parameters such as inductance, capacitance and resistance of the flare loop and oscillations of the electric current in it \citep{Zaitsev1998, Zaitsev2008, Khod2009} varies from several seconds in the solar corona to $10^4$ s in extreme conditions of stellar super-flares. A more detailed discussion of this question would require the forward multi-wavelength modelling and comparative analysis of the expected QPP properties for solar and stellar flare conditions.

\section{Acknowledgments}
This work derives from observations made with the ULTRASPEC instrument at the Thai National Observatory under program ID $TNTC05\_002$ and $TNTC06\_025$, which is operated by the National Astronomical Research Institute of Thailand. Armagh Observatory \& Planetarium is core funded by the N. Ireland Executive through the Dept for Communities. JGD would like to thank the Leverhulme Trust for a Emeritus Fellowship. VSD acknowledges funding from the STFC. NVN is funded via a studentship from AOP. DYK acknowledges support from STFC grant ST/T000252/1. TRM  acknowledges support from STFC grant ST/T000406/1.

\section*{Data availability}
The reduced data underlying this article will be shared on reasonable request to the corresponding author; potential users should 
first contact P. Irawati and V.S Dhillon.

\bibliographystyle{mnras}
\bibliography{ref}
\end{document}